\documentclass{aa}
\usepackage{txfonts,graphicx,subfigure,natbib}
\bibpunct{(}{)}{;}{a}{}{,}
\begin{document}

\title{The galaxies in the field of the nearby GRB\,980425/SN\,1998bw}
\author{S.~Foley\inst{1}
   \and D.~Watson\inst{2}
   \and J.~Gorosabel\inst{3}
   \and J.~P.~U.~Fynbo\inst{2}
   \and J.~Sollerman\inst{2,4}
   \and S.~McGlynn\inst{1}
   \and B.~McBreen\inst{1}
   \and J.~Hjorth\inst{2}}
\institute{Department of Experimental Physics, University College Dublin, Dublin 4, Ireland
      \and Dark Cosmology Centre, Niels Bohr Institute, University of Copenhagen, Juliane Maries Vej 30, DK-2100 Copenhagen \O, Denmark
      \and Instituto de Astrof\'{\i}sica de Andaluc\'{\i}a (IAA-CSIC),
      Apartado de Correos, 3004, 18080 Granada, Spain
      \and Stockholm Observatory, AlbaNova, Department of Astronomy,
      106 91 Stockholm, Sweden}
\date{Received/Accepted}

\abstract{We present spectroscopic observations of ESO\,184$-$G82, the host
          galaxy of GRB\,980425/SN\,1998bw, and six galaxies in its field. A
	  host redshift of $z=0.0087\pm0.0006$ is derived, consistent with
	  that measured by \citet{tinney:1998}. Redshifts are
	  obtained for the six surrounding galaxies
	  observed. Three of these galaxies lie within 11\,Mpc of each
	  other, confirming the suggestion that some of these galaxies form
	  a group. However, all of the field galaxies observed lie at
	  significantly greater distances than ESO\,184$-$G82 and are
	  therefore not associated with it. The host galaxy of
	  GRB\,980425/SN\,1998bw thus appears to be an isolated dwarf
          galaxy and interactions with other
	  galaxies do not seem to be responsible for its star formation.

}

\maketitle

\section{Introduction}

Gamma-Ray Burst GRB\,980425  was
detected by \emph{BeppoSAX} \citep{soffitta:1998} and with the Burst and Transient Source Experiment
\citep[BATSE,][]{kippen:1998}. It is  most
notable for its co-incidence in both space and time with SN\,1998bw
\citep{galama:1998}, a
very energetic Type Ic supernova \citep{patat:2001}. The 
probability of a chance co-incidence of GRB\,980425 and SN\,1998bw has 
been estimated to be $\sim10^{-4}$ \citep{galama:1998}.  
Therefore this detection
provided the first convincing evidence for the link between
long-duration GRBs and SNe,
one which has subsequently been confirmed with the direct detection of
SN\,2003dh associated with GRB\,030329
\citep{hjorth:2003,stanek:2003}. 
This case lent particular weight 
to the proposed GRB\,980425/SN\,1998bw connection due to the spectral 
similarity observed between SN\,1998bw and SN\,2003dh. Additional 
spectroscopic and photometric studies strongly support the GRB-SN
connection \citep{dellavalle:2005,matheson:2005,fynbo:2004}. Examples include the unambiguous association of SN\,2003lw with
GRB\,031203 \citep{malesani:2004,thomsen:2004} and evidence for extra light at
later times in a large proportion of GRB afterglows \citep{zeh:2004}.
 
Assuming this
association to be true, GRB\,980425 was
a sub-luminous GRB, emitting an equivalent isotropic $\gamma$-ray energy of
$\sim$8$\times$10$^{47}$\,erg \citep{galama:1998}, less luminous 
than typical GRBs by a factor of $10^{4}$. It occurred in a spiral arm of
ESO\,184$-$G82, at a redshift of $0.0085\pm0.0002$
\citep{tinney:1998}. 
This makes GRB\,980425 by far the closest GRB detected to
date, the next closest being GRB\,031203 at a redshift of $z=0.1055$
\citep{prochaska:2004}. Given such relative proximity,
GRB\,980425 provides a rare and excellent opportunity for further study of
the host's galaxy field.

Hubble Space Telescope (HST) imaging of ESO\,184$-$G82, as undertaken by
\citet{fynbo:2000}, indicates that the host has a high specific
star-formation rate. This has recently been confirmed by
\citet{sollerman:2005}. Attention is drawn to the local environment of
ESO\,184$-$G82, showing a number of galaxies lying in apparent close
proximity to the GRB host (see Fig.~\ref{figure:1}). ESO\,184$-$G82
has been
considered to be a member of this group of galaxies including ESO\,184$-$ G80 and
ESO\,184$-$G81 \citep{holmberg:1977}. If these galaxies lie at
distances comparable to that of ESO\,184$-$G82, the star-formation rate of
the host galaxy, and hence the probability of a GRB event, may have been
enhanced by interaction with these galaxies.

In this paper we present optical spectroscopy of ESO\,184$-$G82 and
six field galaxies in its vicinity. The paper is organised as follows: Observations
and data reduction procedures are described in Sect.~\ref{observations}.
Section~\ref{analysis} presents the results of our investigation into the
host's environs and Sect.~\ref{discussion} discusses the
conclusions and implications of our results.

\section{Observations and Data Reduction\label{observations}}

\begin{figure}
\resizebox{\hsize}{!}{\includegraphics{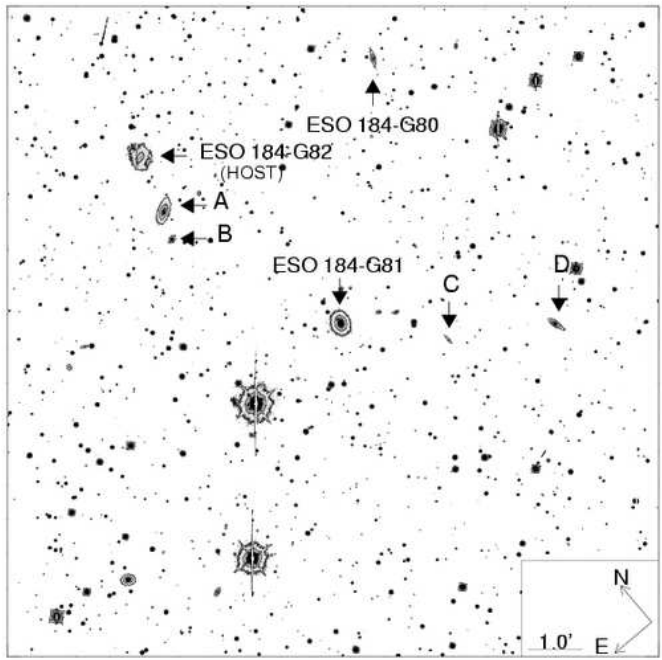}}
\caption{DFOSC R-band field surrounding GRB\,980425
  (13.7\arcmin\,$\times$13.7\arcmin\ FOV). Galaxies 
for which spectroscopic observations were carried out
 are marked (see Table~\ref{table:1}).} 
\label{figure:1}
\end{figure}

Spectroscopic observations were undertaken of the host and six 
galaxies in
its field, as marked in Fig.~\ref{figure:1}. The field 
galaxies include ESO\,184$-$G80, ESO\,184$-$G81 and four 
previously unclassified galaxies, labelled A to D with increasing 
angular distance from the host.

The field  galaxies were observed using the Danish 1.54\,m
telescope at La Silla, Chile. The Danish Faint Object Spectrograph and 
Camera (DFOSC) was mounted on the telescope with grism \#\,7 for all 
observations, resulting in a spectral range of 3800--6800\,\AA. The
slit width was varied between 2.5\arcsec{} and 2.0\arcsec{} giving a
spectral resolution of $\sim$10\,\AA\,and
$\sim$8\,\AA,\,respectively, measured from the typical FWHM of an arc
spectrum emission line. 
In most cases, the slit was aligned so as to include two of the sample 
galaxies.  The detector used was a 2k\,$\times$\,2k Loral CCD with a 
gain of 1.3\,e$^-$\,ADU$^{-1}$, read-out noise of 7.5\,e$^-$ and 
pixel scale of 0.40\arcsec{} pixel$^{-1}$. The DFOSC field of view
(FOV) is 
13.7\arcmin\,$\times$13.7\arcmin\,. In order to measure the redshift of the host galaxy we
used the VLT spectra obtained on June 13 1999 by
\citet{sollerman:2000}. These observations were carried out using the FORS1 instrument. Grism 300V was used, giving a
wavelength range of 3600--9000\,\AA. The width of the slit was 1\arcsec{}
giving a spectral resolution of $\sim$7\,\AA. The gain and read-out noise of the
CCD were set to 0.34\,e$^-$\,ADU$^{-1}$ and 5.75\,e$^-$,
respectively. These data are also available in the ESO archive.
 
Zero-exposure bias frames and flat-field exposures were taken on each 
observing night. Spectral images of He-Ne arc lines were obtained 
immediately before and subsequent to the object exposures under 
identical observing conditions for the Danish 1.54\,m
observations. For the stable VLT/FORS1 instrument, we used arc
images obtained in the morning as part of the usual calibration
plan. For the VLT/FORS1 observations of the host galaxy, we also
performed observations of the standard star LTT\,7379. Spectroscopic 
standard star observations were not obtained 
for the field galaxies. Thus, the DFOSC spectra are not calibrated in flux (see panels
(b-g) of Fig.~\ref{figure:2}), which is irrelevant for our purposes.
A log of the observations is shown in Table~\ref{table:1}.

\begin{table*}
\caption{Log of observations.}
\label{table:1}
\begin{tabular*}{2\columnwidth}{@{\extracolsep{\fill}}@{}c c c c c@{}}
\hline 
Telescope/Grism & Object & Date & Exposure Time (s) & Seeing
  (\arcsec) \\
\hline \hline 
VLT/300V       & ESO184$-$G82  & 13-06-1999 & 4$\times$1800  & \\
D\,1.54m/\#\,7 & ESO 184$-$G80 & 24-08-2000 & 3$\times$1200, 1$\times$1050 & 1.8 \\
              % &               &            & 1$\times$1050  & \\
               &               & 30-08-2000 & 6$\times$1200  & 1.2 \\
               & ESO 184$-$G81 & 27-08-2000 & 5$\times$1200  & 1.4 \\
               &               & 28-08-2000 & 7$\times$1200  & 1.6 \\
               &               & 01-09-2000 & 4$\times$1200  & 1.4 \\
               & Galaxy A      & 26-08-2000 & 2$\times$1200, 1$\times$1118 & 2.3 \\
              % &               &            & 1$\times$1118  & \\
               &               & 29-08-2000 & 6$\times$1200  & 1.6 \\
               &               & 02-09-2000 & 5$\times$1200  & 1.0 \\
               & Galaxy B      & 26-08-2000 & 2$\times$1200, 1$\times$1118 & 2.3 \\
              % &               &            & 1$\times$1118  & \\
               &               & 29-08-2000 & 6$\times$1200  & 1.6 \\
               &               & 02-09-2000 & 5$\times$1200  & 1.0 \\
               & Galaxy C      & 27-08-2000 & 5$\times$1200  & 1.4 \\
               &               & 30-08-2000 & 6$\times$1200  & 1.2 \\
               & Galaxy D      & 28-08-2000 & 7$\times$1200  & 1.6 \\
               &               & 01-09-2000 & 4$\times$1200  & 1.4 \\
\hline 
\end{tabular*}
\end{table*}

Data reduction was performed as standard using 
IRAF~V2.12.2\footnote{Image Reduction and Analysis Facility (IRAF) 
is distributed by the National Optical Astronomy Observatories
(NOAO).}. Frames were bias and overscan-subtracted in order to remove 
the pedestal bias level and any underlying pixel structure. Flat-field 
frames were divided into the object and arc frames to eliminate 
pixel-to-pixel sensitivity variations. The object spectra were then 
extracted to 1-D images of pixel number against ADU pixel$^{-1}$. A 
variance-weighted extraction was chosen to remove random cosmic ray
events. In order to wavelength calibrate the pixel scale, calibration 
spectra were extracted from the arc lamp images and a dispersion scale 
determined from the known wavelengths of a number of Hg-He-Ar/He-Ne 
emission lines. Dispersion scales were assigned to object spectra by 
taking an average of those arc spectra solutions from the calibration 
exposures nearest in time to the object exposures. The typical rms of 
the dispersion fit ranged from 0.01\,\AA\ to 0.1\,\AA. The accuracy
of the wavelength calibration was verified by identifying the 
wavelengths of prominent sky lines for each night. The observed 
deviations are consistent with the above rms scatter. No systematic 
variations in the wavelength scale from night to night were observed. 
 A flux calibration was performed for the host galaxy using the 
observed flux of the spectrophotometric standard LTT\,7379.    

\section{Analysis\label{analysis}}

The redshift of each galaxy was determined using two methods 
(see Table~\ref{table:2}). An 
initial determination of the redshift was made using the IRAF task 
\emph{rvidlines}. This requires the identification of a prominent
spectral feature to which a gaussian function is fit. Based on the 
central wavelength of this line and an input list of known spectral 
lines, other features at a consistent redshift are identified. A 
weighted average redshift value and corresponding error based on the 
gaussian fit is output.

An alternative measurement of each redshift was obtained using the 
Fourier cross-correlation method as described by \citet{tonry:1979} using the IRAF task \emph{fxcor}.  In this analysis, 
each galaxy spectrum was cross-correlated with template spectra 
obtained from \citet{kinney:1996}. These spectra are 
characterised by high S/N ratios and are de-redshifted to the rest 
frame. Template spectra for bulge, elliptical, S0, Sa, Sb 
(absorption-line dominated) and 
Sc (emission-line dominated) galaxy morphologies were used. \emph{Fxcor}
determines the wavelength shift of the object spectrum at the point of  
maximum correlation and computes a redshift value. The task outputs
redshift errors based on the height of the correlation peak and noise 
statistics.  

\begin{table}
\caption{Heliocentric-corrected redshifts for galaxies in the field of
GRB\,980425 as marked in Fig.~\ref{figure:1}}
\label{table:2}
\begin{tabular}{@{}c c c@{}}
\hline
Galaxy & Redshift (\emph{fxcor}) & Redshift (\emph{rvidlines}) \\
\hline \hline
ESO 184$-$G82 & $0.0085\pm0.0003$ & $0.00867\pm0.00004$ \\
ESO 184$-$G80 & $0.0470\pm0.0007$ & $0.0470\pm0.0002$ \\
ESO 184$-$G81 & $0.0448\pm0.0007$ & $0.0452\pm0.0001$ \\
Galaxy A & $0.0447\pm0.0007$ & $0.0448\pm0.0003$ \\
Galaxy B & $0.0314\pm0.0003$ & $0.0317\pm0.00002$ \\
Galaxy C & undetermined & $0.0609\pm0.0001$ \\
Galaxy D & $0.0605\pm0.0006$ & $0.0604\pm0.0002$ \\
\hline
\end{tabular}
\end{table}  

\begin{figure*}
  \begin{minipage}[t]{.5\textwidth}
    \vspace{0pt}
    \centering
    \includegraphics[bb=0 0 504 1008,width=\columnwidth,height=15cm,clip=]{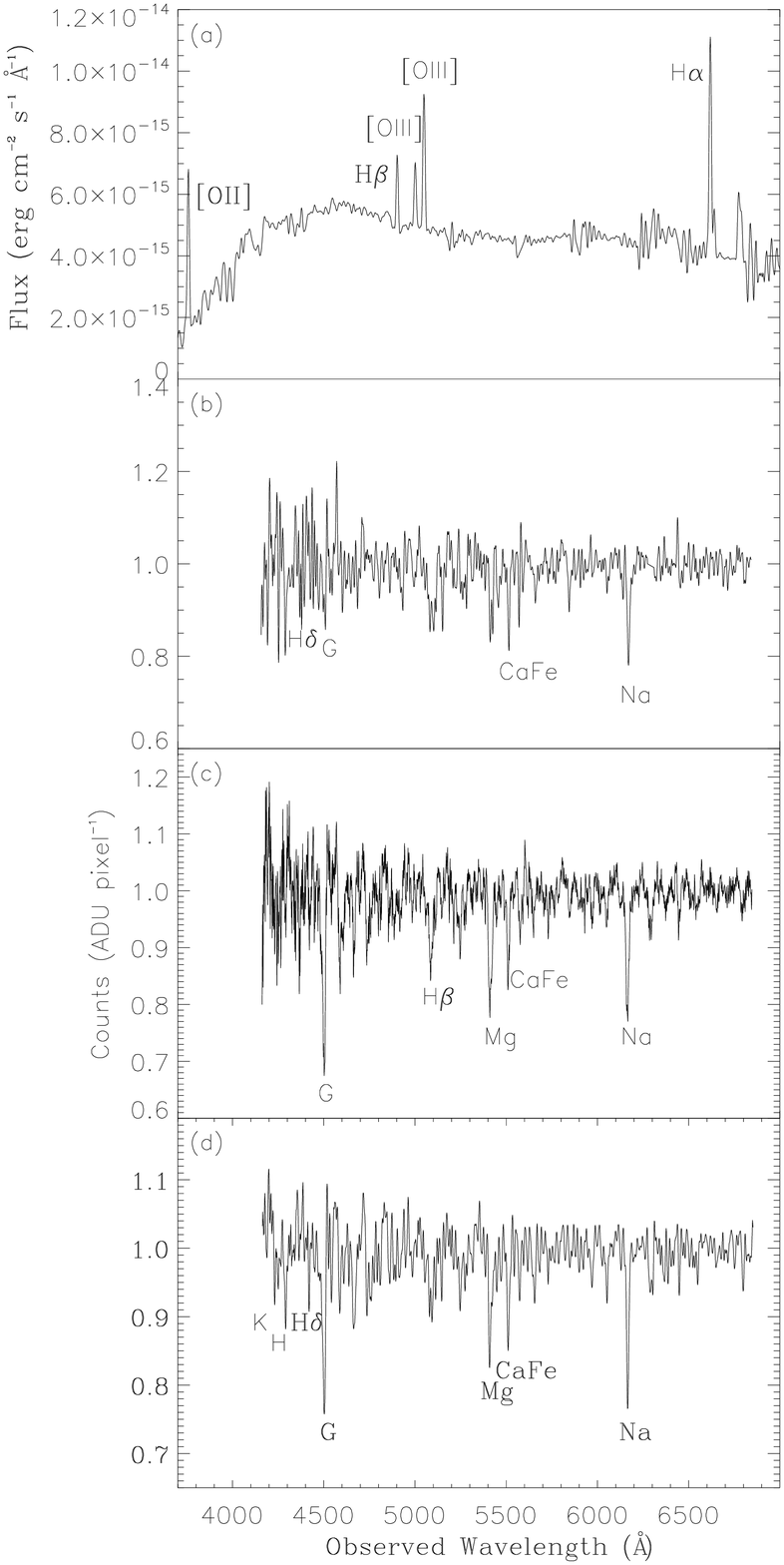}
  \end{minipage}%
  \begin{minipage}[t]{.5\textwidth}
    \vspace{0pt}
    \centering
    \includegraphics[bb=0 0 504 777,width=\columnwidth, height=15cm,clip=]{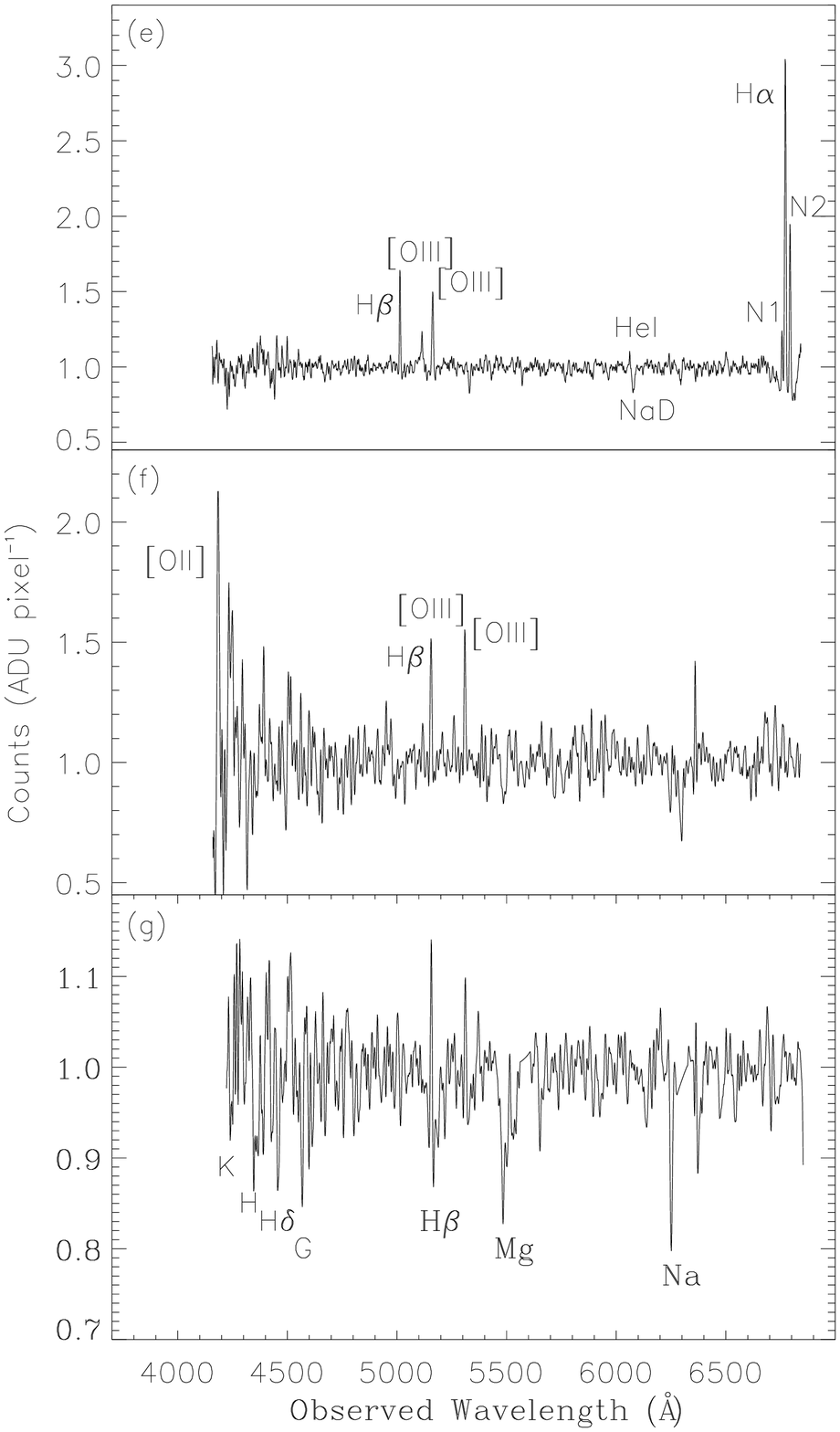}
  \end{minipage}
\caption{Spectra of (a)\,ESO\,184$-$G82, the host of GRB\,980425 and
 galaxies in its field:
  (b)\,ESO\,184$-$G80, (c)\,ESO\,184$-$G81, (d)\,Galaxy A, (e)\,Galaxy B,
  (f)\,Galaxy C and (g)\,Galaxy D. The VLT spectrum of the host (a) is
 flux-calibrated while the DFOSC spectra (b-g) are
 normalised to a flat continuum. The spectral lines used in 
the redshift determination for each galaxy are labelled.} 
\label{figure:2}
\end{figure*}

Combined spectra for each galaxy observed are shown in
Fig.~\ref{figure:2}. Table~\ref{table:2} gives the heliocentric 
redshifts determined in each case using the \emph{fxcor} and 
\emph{rvidlines} methods discussed above. Through the use of a 
number of templates for correlation with the absorption spectra, a 
mean value for the redshift was determined for a given galaxy
spectrum, weighted by the square inverse of the \emph{fxcor}
error. The \emph{fxcor} errors from each template are 
added in quadrature to produce the \emph{fxcor} error quoted in 
Table~\ref{table:2}. A 
consistent value of $\sigma=0.0004$ for the standard deviations of the
independent cross-correlations was obtained for each galaxy. 
The spectrum obtained for
Galaxy C proved too weak for an accurate \emph{fxcor} redshift 
determination. 
 
 In the case of an 
emission-line dominated spectrum, the 
\emph{rvidlines} redshift is adopted, since the emission features
are unambiguous and well fit by a gaussian function. For the 
absorption-line spectra, we view the \emph{fxcor} redshift as being
the more reliable method, as it proved difficult to fit gaussians to
the absorption lines. In such cases \emph{rvidlines} was used to verify 
the authenticity of the chosen correlation peak. The \emph{rvidlines} error in Table~ \ref{table:2} is an internal 
error generated by the task based solely on the gaussian fit to a single
spectral feature and so is significantly underestimated. We therefore assume a conservative
error of 0.0006 for all redshifts based on the template redshift deviations
and the \emph{fxcor} errors. 

The redshift difference between the host galaxy and the bright galaxies in
its field corresponds to a velocity difference of 6000\,km\,s$^{-1}$, while the typical velocity 
dispersion observed in low redshift galaxy clusters is at most 1000\,km\,s$^{-1}$
\citep{desai:2004}. 
We therefore conclude that
ESO\,184$-$G82 is an isolated dwarf galaxy in a state of active
star-formation, the origin of which is not attributable to any obvious
interaction with its local environment. Having a velocity dispersion of
$\sim300$ km\,s$^{-1}$ and lying within 11\,Mpc of each other, 
ESO\,184$-$G81, ESO\,184$-$G80 and Galaxy~A are most
likely located in a small group, but are not near to the host galaxy.
   
\section{Are GRB hosts in underdense regions?\label{discussion}}

In an extensive HST imaging study of 42 GRB host galaxies,
a large proportion of those observed ($\sim$30--60\%) show evidence
for interaction \citep{wainwright:2005}. Analysis of the environments
of four GRB hosts identify several galaxies at a similar redshift in all cases \citep{fynbo:2002,jakobsson:2005}. GRB\,980613 also occurred in a host galaxy
which appears to be part of a complex, interacting system
\citep{hjorth:2002} as is the host of GRB\,001007
\citep{castro:2002}. However, it is not known if any of these fields are
overdense. 

Whether GRBs preferentially occur in galaxy-dense regions remains uncertain,
but the evidence seems to hint that they do not. Photometric redshifts have
been determined for galaxies in the field of GRB\,000210 showing that there
is no obvious grouping of galaxies around the host \citep{gorosabel:2003}. In a study of the cross-correlation between GRB host
galaxies and the surrounding galaxy environments, 
\citet{bornancini:2004} found tentative evidence that GRB host
galaxies are more
likely to be located in low density galaxy environments. It has been suggested
that objects found in lower-density galaxy regions may also tend
to be sub-luminous \citep{lefloch:2003}, consistent with
ESO\,184$-$G82 in particular and GRB host characteristics in general. The 
GRB\,030329 host at z=0.1685 \citep{greiner:2003} may be a counter-example, having at least one field 
galaxy at $z\sim$0.171 and hence being consistent with a potential clustering 
\citep{gorosabel:2005}.

We can now conclude that ESO\,184$-$G82, generally regarded as being part of
a galaxy group, appears in fact to be isolated and that interaction with nearby
galaxies can be eliminated as a source of its high specific star-formation
rate. Interestingly, in a recent study, \citet{sollerman:2005}
find that the host galaxy properties are consistent with a constant
star-formation rate over a few Gyrs.  Therefore, it seems that the
present state of the galaxy is not that of a starburst and hence that the
current star-formation activity does not necessarily require an external
trigger.

\section*{Acknowledgements}
The Dark Cosmology Centre is supported by the DNRF. The
authors acknowledge benefits from collaboration within the EC FP5 Research
Training Network "Gamma-Ray Bursts - An Enigma and a Tool". This research is partially
supported by the Spanish Ministry of Science and Education through
programmes ESP2002-04124-C03-01 and AYA2004-01515.

\bibliographystyle{aa}
\bibliography{refs}

\end{document}